\documentclass{aa}  
\usepackage{natbib}
\usepackage{amsmath}
\usepackage{float} 
\bibpunct{(}{)}{;}{a}{}{,} 
\usepackage{graphicx}
\usepackage{txfonts}
\usepackage{lipsum}
\usepackage{subcaption}         
\usepackage{lscape}             
\usepackage{placeins}           
                                
\usepackage{hyperref}
\usepackage{siunitx}
\usepackage{caption}
\usepackage{natbib}   
\usepackage{subcaption}
\hypersetup{
    colorlinks=true,
    linkcolor=blue,
    urlcolor=blue,
    citecolor=blue
    }

\newcommand{\equalcontrib}{\textsuperscript{\textdagger}}
\RequirePackage{etex} 
\begin{document}

   \title{Collisional excitation of $E$- and $Z$-ethanimine to model cold molecular clouds}


   \author{Francesca Tonolo \inst{1}\equalcontrib
        \and Vivek Vijay\inst{2}\equalcontrib
        \and Ernesto Quintas-Sánchez \inst{3}\equalcontrib       
        \and Adrian Batista Planas\inst{3}
        \and Carolin Joy \inst{2} 
        \and Richard Dawes \inst{3} 
        \and François Lique\inst{1}\fnmsep\thanks{Email: francois.lique@univ-rennes.fr}
        \and Dmitri Babikov\inst{2}\fnmsep\thanks{Email: dmitri.babikov@marquette.edu}
        }

   \institute{Univ Rennes, CNRS, IPR (Institut de Physique de Rennes) - UMR 6251, F-35000 Rennes, France.
   \and Chemistry Department, Wehr Chemistry Building, Marquette University, Milwaukee, WI 53201-1881, USA.
   \and Department of Chemistry, Missouri University of Science and Technology, Rolla, Missouri 65409, USA.}

   \date{Accepted \today}

 
  \abstract
   {Ethanimine is a prebiotically relevant complex organic molecule that has been proposed as a key precursor of amino acids in the interstellar medium. To date, both its $E$ and $Z$ isomers have only been detected toward the Sgr B2(N) molecular cloud. The pronounced non-local thermodynamic equilibrium (non-LTE) effects predicted for this molecule require accurate collisional rate coefficients for reliable astrophysical modeling and future astronomical searches.}
   {We aim to provide the first collisional datasets for both ethanimine isomers with the main cold molecular clouds collision partners, He and \emph{para}-H$_2$, in order to support non-LTE radiative-transfer modeling.}
   {State-to-state rate coefficients for collisions with He were computed by integration over mixed close-coupling (CC) and coupled-states (CS) scattering cross sections at different energy ranges. In addition, comparisons with mixed quantum-classical theory (MQCT) calculations performed for both He and \emph{para}-H$_2$ collisions enabled the derivation of transition-specific scaling factors, which were applied to predict collisional rate coefficients for \emph{para}-H$_2$.}
   {We present the first state-to-state collisional datasets for both $E$- and $Z$-ethanimine in collision with He and \emph{para}-H$_2$, including all rotational transitions involving levels below 50\,cm$^{-1}$ and over the 5--100\,K temperature range.} 
   {The datasets reported here provide the first collisional information suitable for non-LTE modeling of ethanimine in cold interstellar environments. They represent a key step toward more reliable abundance determinations and may facilitate future astronomical detections of both ethanimine isomers beyond Sgr B2(N).} 


   \maketitle

\begingroup
\renewcommand{\thefootnote}{\textdagger}
\footnotetext{These authors contributed equally to this work.}
\endgroup
\section{Introduction}
\nolinenumbers

Ethanimine (CH$_3$CHNH) is an interstellar complex organic molecule (COM) of astrochemical and prebiotic relevance, as it has been proposed as a key intermediate in the formation of biologically important species, such as amino acids, in the interstellar medium (ISM; \cite{quan2016chemical}). 
As an example, previous studies have identified ethanimine as one of the main precursors of alanine (C$_3$H$_7$NO$_2$) through reactions involving HCN and H$_2$O, or alternatively formic acid (HCOOH; \cite{woon2002pathways,elsila2007mechanisms,loomis2013detection}).

This molecule exists in two geometrical isomeric forms, $E$- and $Z$-ethanimine, which differ in the orientation of the hydrogen atom within the -NH moiety, separated by an energy difference of approximately 552\,K, with the $E$-ethanimine being the most stable isomer \citep{quan2016chemical}.

Both $E$- and $Z$-ethanimine isomers have been detected toward the Sgr B2(N) molecular cloud as part of the Green Bank Telescope PRebiotic Interstellar MOlecule Survey (GBT PRIMOS; \cite{loomis2013detection}). 
However, state-of-the-art quantum-chemical and kinetic models have so far failed to reproduce the observed $E$/$Z$ isomeric abundance ratio. Indeed, values of \emph{ca.}\,1.2 and 1.4 were predicted by \cite{balucani2018theoretical} and \cite{baiano2020role}, respectively, consistently underestimating the value of \emph{ca.}\,3 inferred from astronomical observations \citep{loomis2013detection}. 
Furthermore, despite the extensive experimental and theoretical characterization of the rotational and infrared spectra of ethanimine (see \cite{melli2018rotational} and references therein), no additional astronomical detection of this species has yet been confirmed. 

A possible explanation is that the search for ethanimine and the astrophysical modeling of its emission lines may have been hindered by the complete lack of collisional data with the most abundant interstellar collision partners, namely H$_2$, He, and H.
Indeed, the strong departures from local thermodynamic equilibrium (LTE) that characterize many astrophysical environments require to account for the delicate balance between radiative and collisional processes in determining molecular level populations \citep{roueff2013molecular}. Accurate astrophysical modeling therefore requires reliable state-to-state collisional information.
\begin{figure*}[h!]
   \centering
   \includegraphics[scale=0.27]{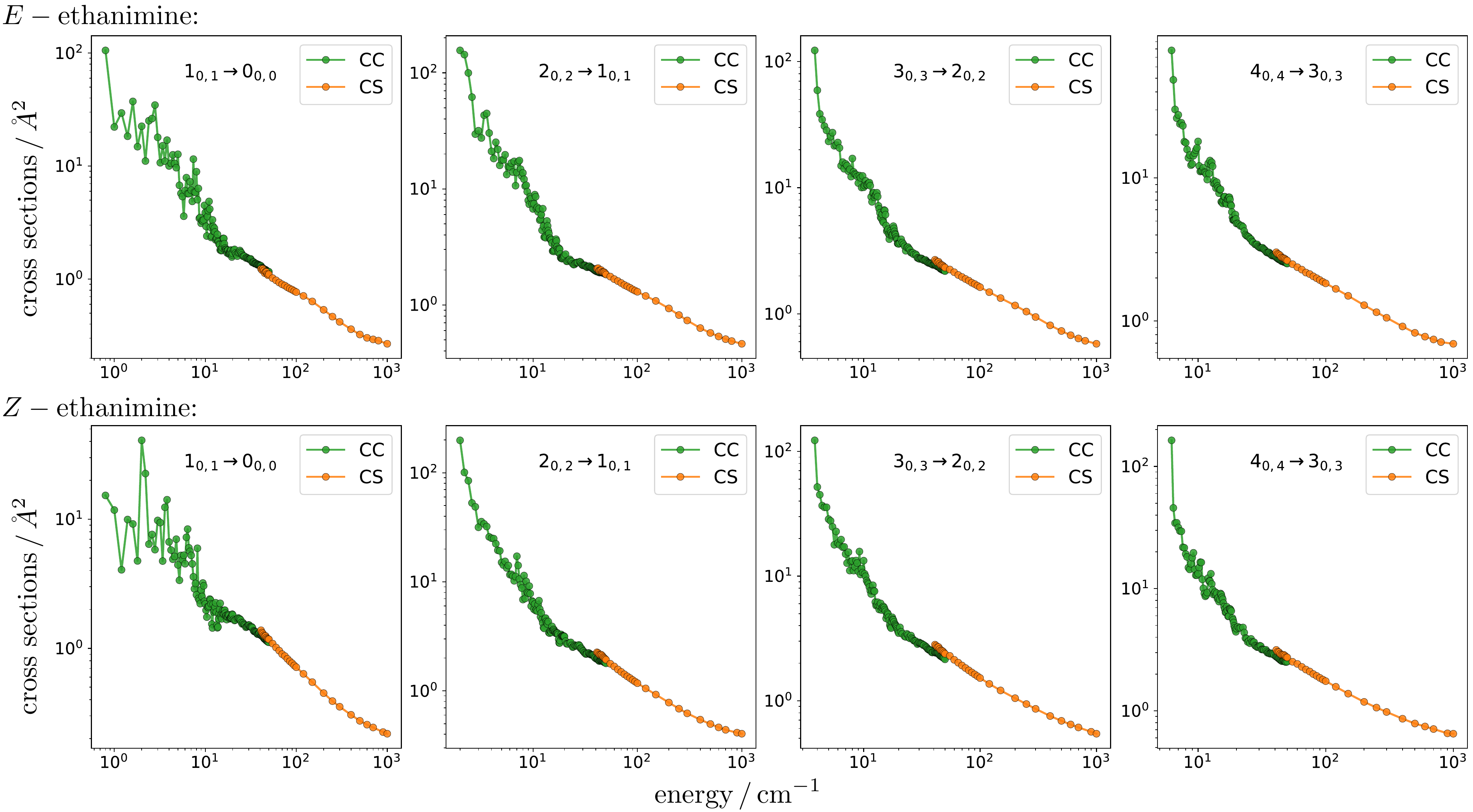}
      \caption{Inelastic cross sections of the $E$ and $Z$ isomers of ethanimine in collision with He for the lowest $a$-type de-excitation rotational transitions. The cross sections in green were obtained by means of full quantum CC calculations up to 50\,cm$^{-1}$, while the cross sections in orange refer to CS calculations in the 40--1000\,cm$^{-1}$ energy range.}
         \label{fig1}
\end{figure*}

Recently, an astrophysical modeling of ethanimine observations has been conducted by including approximate collisional rate coefficients obtained through scaling relations \citep{sharma2023potential}. The latter revealed the presence of several weak maser transitions and anomalous absorption features for both isomers. These findings suggest that accurate collisional datasets may play a crucial role in guiding new identifications of ethanimine in a wider variety of astrophysical environments.
 
To fill this gap, we computed the first state-to-state collisional datasets for both ethanimine isomers in collision with He using accurate quantum-scattering methods. The collisional rate coefficients were derived from a combined set of inelastic cross sections obtained from full quantum close-coupling (CC; \cite{arthurs1960theory}) and coupled-states (CS; \cite{mcguire1974quantum}) calculations, whose derivation was previously reported by the authors in \cite{vijay2026collisional} (hereafter Paper~I). 

We subsequently applied a transition-dependent scaling approach, based on mixed quantum-classical theory (MQCT; \cite{semenov2015mixed,mandal2024rotational,bostan2024mixed,joy2025mixed}) calculations, to derive the predictions of these coefficients for collisions with \emph{para}-H$_2$. These data provide a reliable basis for non-LTE radiative-transfer modeling and may facilitate future astronomical detections of both $E$- and $Z$-ethanimine.

This paper is organized as follows. Section~\ref{sec2} presents the methodology adopted to derive the state-to-state collisional rate coefficients. Section~\ref{sec3} reports the results and discusses the features most relevant to astrophysical modeling, including the application of the transition-specific scaling factors. Finally, in Sect.~\ref{sec4} the main conclusions are drawn. 

\section{Methodology}
\label{sec2}
\begin{figure*}[h!]
   \centering
   \includegraphics[scale=0.25]{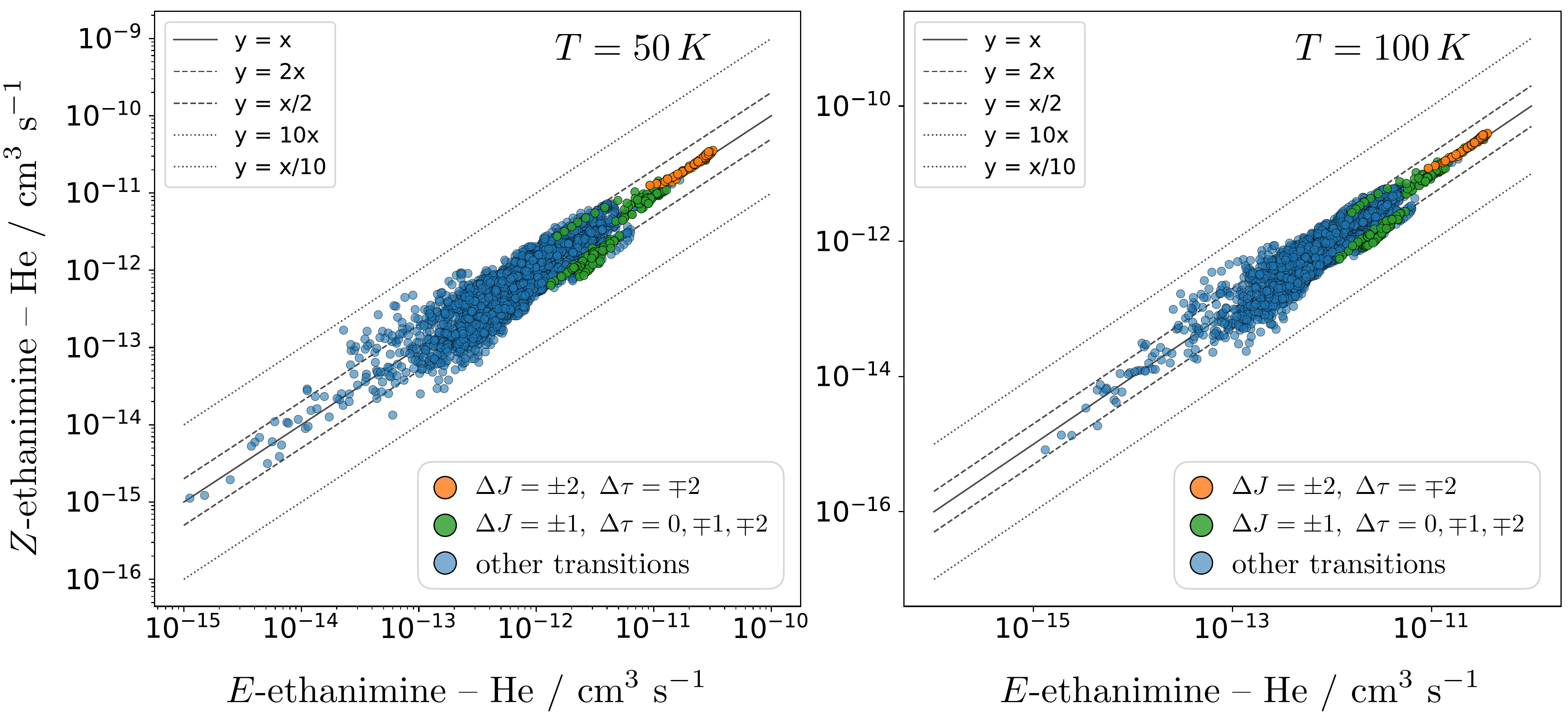}
      \caption{Comparison of the inelastic collisional rate coefficients between the $E$ and $Z$ isomers of ethanimine in collision with He, at 50 and 100\,K. Different colors highlight the main collisional propensities: orange symbols correspond to transitions with $\Delta J=\pm2$ and $\Delta\tau=\mp2$; green symbols to transitions with $\Delta J=\pm1$ and $\Delta\tau=0,\mp1,\mp2$; blue symbols represent all remaining transitions.}
         \label{fig2}
\end{figure*}
The state-to-state collisional rate coefficients of $E$- and $Z$-ethanimine by He were obtained by thermally averaging the corresponding inelastic cross sections retrieved in Paper~I over the collision energy. To ensure a reliable convergence of the thermal average over the 5--100\,K temperature range, which is relevant to cold molecular clouds, the energy grid must extend sufficiently beyond the energies contributing to the Maxwell--Boltzmann distribution. However, owing to the high computational cost of full quantum CC calculations, the CC cross sections were only computed up to $E_{\text{tot}} = 50$\,cm$^{-1}$, which is insufficient to guarantee a robust convergence of the thermal integration.
Nevertheless, as illustrated in Fig.~\ref{fig1} for selected transitions of both ethanimine isomers, the CS calculations were able to reproduce the CC cross sections with very good agreement over the intermediate energy range. In particular, the two datasets showed a good overlap between 40 and 50\,cm$^{-1}$, with average relative deviations always less than 20\%. This provided the basis for constructing a combined set of inelastic cross sections, consisting of CC results for $E_{\text{tot}}\leq 50$\,cm$^{-1}$, and CS results for $E_{\text{tot}}\geq 50$\,cm$^{-1}$. 
The rate coefficients were then obtained by thermally averaging the combined set of cross sections. Prior to the integration, a spline interpolation was applied to the complete set of cross sections in order to generate a smooth representation as a function of collision energy. This procedure also ensured a continuous transition between the CC and CS energy ranges, avoiding artificial discontinuities at the matching energy.

Following this procedure, we computed the first set of rotational (de-)excitation rate coefficients for both $E$- and $Z$-ethanimine isomers in collision with He over the 5--100\,K temperature range. The datasets include all transitions involving the rotational levels with energies below 50\,cm$^{-1}$, corresponding to the levels up to $6_{5,2}$\footnote{
In the following, the $J_{K_a,K_c}$ notation for an asymmetric rigid rotor is adopted to label the rotational levels of $E$- and $Z$-ethanimine. Here, $J$ denotes the rotational angular momentum quantum number, while $K_a$ and $K_c$ are pseudo-quantum numbers that correlate with the symmetric-top quantum number $K$ in the prolate and oblate limits, respectively. 
} for $E$-ethanimine and $9_{4,6}$ for $Z$-ethanimine. 
Higher-lying rotational levels were not considered because they are more strongly coupled to states lying outside the rotational basis employed in the scattering calculations, preventing the corresponding cross sections and rate coefficients from reaching satisfactory convergence.
It should be noted that above 100\,K, in absence of strongly subthermal conditions, higher-lying rotational states become increasingly populated and contribute to the excitation balance. Therefore, extrapolation of the present dataset to higher temperatures should be treated with caution.

\section{Results and Discussion}
\label{sec3}
\begin{figure*}[h!]
   \centering
   \includegraphics[scale=0.40]{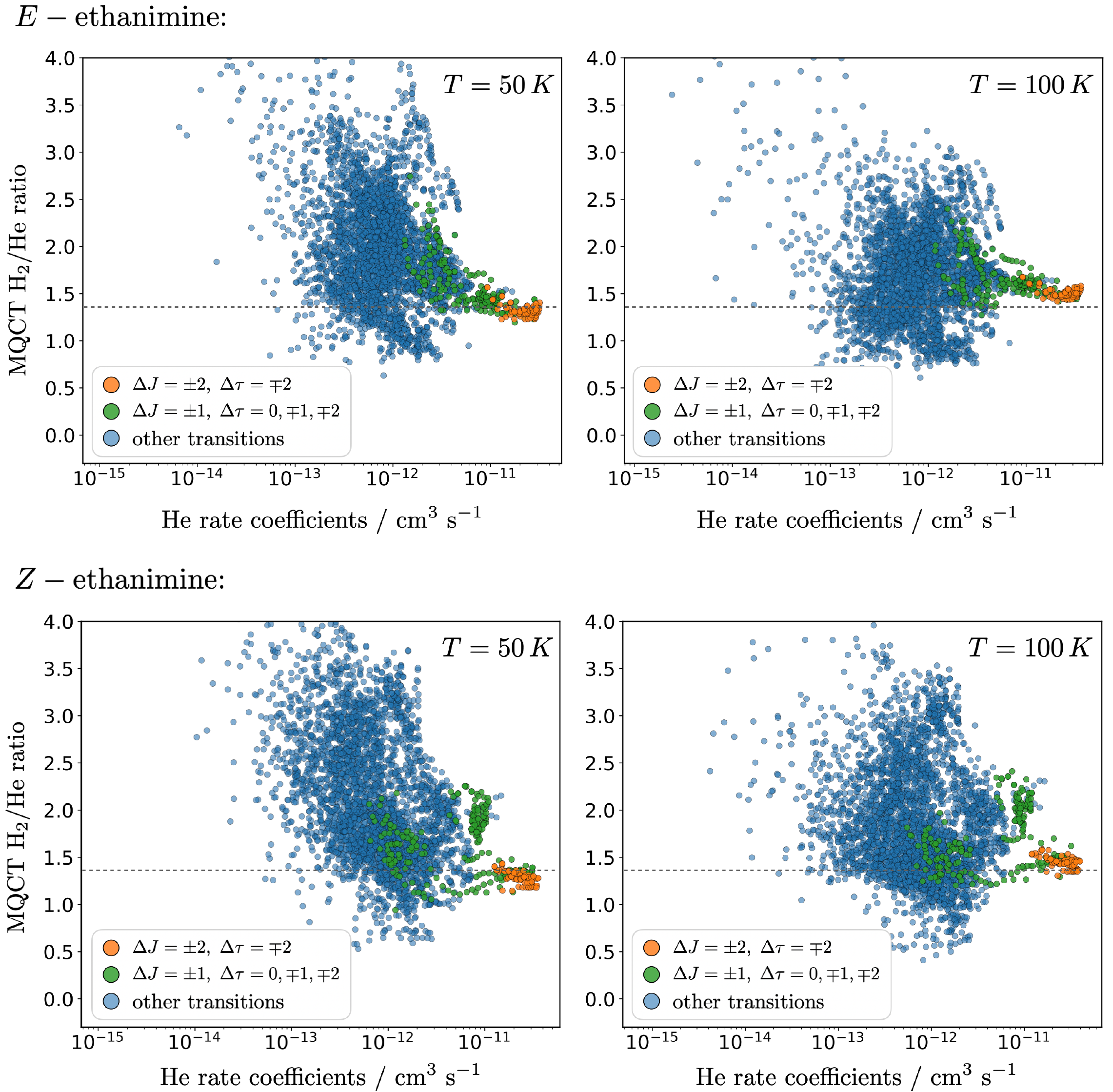}
      \caption{Ratio between the de-excitation collisional rate coefficients by \emph{para}-H$_2$ and He as a function of the corresponding He rate coefficient, obtained from MQCT calculations, for the $E$ and $Z$ isomers of ethanimine at 50 and 100\,K. The horizontal dotted line indicates the reduced-mass scaling factor (1.4) commonly adopted to estimate H$_2$ rate coefficients from He data. Different colors highlight the main collisional propensities: orange symbols correspond to transitions with $\Delta J=\pm2$ and $\Delta\tau=\mp2$; green symbols to transitions with $\Delta J=\pm1$ and $\Delta\tau=0,\mp1,\mp2$; blue symbols represent all remaining transitions.}
         \label{fig3}
\end{figure*}
Figure~\ref{fig2} compares the collisional rate coefficients of the two isomers of ethanimine at 50 and 100\,K. Overall, the collision dynamics show only a weak dependence on both temperature and isomeric form, with the closest agreement found for the largest rate coefficients, which are expected to dominate the excitation balance under astrophysical conditions.
For both isomers, the strongest propensity is observed for transitions satisfying $\Delta J=\pm2$ and $\Delta\tau=\mp2$ (highlighted in orange in Figure~\ref{fig2}), where $\tau= K_a - K_c$. 
In terms of the asymmetric-top quantum numbers, this corresponds predominantly to $a$-type transitions (with $\Delta K_a=0$) and $\Delta K_c=\pm 2$.
A second, less pronounced propensity favors transitions with $\Delta J=\pm1$ and $\Delta\tau=0,\mp1,\mp2$ (shown in green in Figure~\ref{fig2}), which corresponds primarily to $a$-type transitions with $\Delta K_c=0,\pm 1, \pm 2$. These results are consistent with the propensity rules identified from the inelastic cross sections presented in Paper~I.

Subsequently, we investigated whether these datasets could be extended to collisions with \emph{para}-H$_2$, the dominant collider in the interstellar medium. Performing full quantum calculations for the H$_2$ system is substantially more demanding than for He because the deeper interaction potential requires a much larger number of scattering channels to be included in the calculations, significantly impacting on the computational cost. We therefore explored an alternative strategy based on transition-specific scaling factors derived from the computationally affordable MQCT approach. As demonstrated in Paper~I for the ethanimine-He system, MQCT reproduces the quantum calculations remarkably well at intermediate and high collision energies, while reducing the computational cost by orders of magnitude.
Hence, we computed new state-of-the-art interaction potentials for both the $E$- and $Z$-ethanimine isomers interacting with \emph{para}-H$_2$, and performed MQCT scattering calculations considering H$_2$ in its lowest rotational state. The construction of the interaction potentials and the MQCT calculations are described in Appendix~A. Figure~\ref{fig3} reports the resulting H$_2$/He ratios of the collisional rate coefficients at 50 and 100\,K.
For the dominant transitions, the MQCT H$_2$/He ratio converges to a value close to $\sim1.4$, with average values that range from 1.33 to 1.51 depending on the temperature, and maximum values of 1.57/1.67.
This was somewhat expected, since this factor corresponds to the difference in reduced mass between the two colliders, and is usually employed in astrophysical modeling to scale the He rate coefficients when collisional coefficients with H$_2$ are not available.
However, although weaker transitions contribute less to the overall excitation, many of them exhibit H$_2$/He ratios that deviate significantly from this value, with average ratios ranging between 1.76 and 2.00 and maximum values of 7.15 at 100\,K and 16.22 at 50\,K. This demonstrates that a simple reduced-mass scaling cannot accurately reproduce the effect of the H$_2$ collider for all transitions.
Moreover, the H$_2$/He ratio shows a weak but systematic temperature dependence. As the temperature increases, the ratio for the dominant transitions becomes slightly larger than 1.4. A possible explaination is that at higher collision energies 
the shorter lifetime of the intermediate complex enhances the sensitivity of the collision dynamics to the anisotropic components of the interaction potential,
leading to inelastic cross sections that do not scale solely with the reduced mass. 

These results indicate that uniformly scaling the He rate coefficients by the 1.4 reduced-mass factor provides only a first-order approximation. We therefore derived transition-specific corrections from the MQCT calculations and applied them to each collisional rate coefficient obtained with the combined CC+CS approach at the corresponding temperature. For temperatures below 50\,K, where MQCT method becomes less reliable, we adopted the scaling factors computed at 50\,K.

The complete sets of state-to-state collisional rate coefficients 
will be made publicly available through the EMAA \citep{faure2025excitation}, LAMDA \citep{schoier2005atomic}, and BASECOL \citep{dubernet2024basecol2023} databases.


\section{Conclusions}
\label{sec4}
The present study aims to support the interpretation of current and future observations of the $E$ and $Z$ isomers of ethanimine, an important prebiotic molecule, in cold molecular clouds. In these astrophysical environments, the strong departures from LTE conditions require accurate collisional datasets with the most abundant colliders (typically H$_2$, He, and H) to ensure reliable astrophysical modeling. Moreover, the presence of maser emission and anomalous absorption features in ethanimine observations makes such datasets even more crucial for guiding future astronomical studies.

In this work, we computed the first dataset of state-to-state inelastic rate coefficients involving the rotational levels below 50\,cm$^{-1}$ for both $E$- and $Z$-ethanimine in collision with He over the temperature range 5--100\,K. We combined quantum CC and CS results for the inelastic cross sections below and above 50\,cm$^{-1}$, respectively, to derive the state-to-state rate coefficients by thermal averaging over the collision energy.
The resulting collisional rate coefficients exhibit a strong propensity for transitions involving changes in the rotational quantum numbers $J$ and $\tau$ of $\pm2$ and $\pm1$. They also show a weak dependence on both the temperature and isomeric form.
Finally, in order to facilitate the direct application of these datasets in non-LTE radiative transfer models, we extended them to collisions with \emph{para}-H$_2$. This was achieved by applying transition-specific scaling factors derived from MQCT calculations performed for both ethanimine isomers in collision with He and \emph{para}-H$_2$.

The datasets presented in this work will play a central role in determining the non-LTE  distribution of population of the $E$ and $Z$ isomers of ethanimine in cold molecular clouds. They will provide new knowledge to probe the chemical and physical conditions interstellar environments, and will guide future observational searches.

\begin{acknowledgements}
This research was supported by NSF grant number CHE-2102465. This work used Anvil CPU resources at Purdue University through allocation PHY260027 from the Advanced Cyberinfrastructure Coordination Ecosystem: Services \& Support (ACCESS) program, which is supported by National Science Foundation grants n 2138259, 2138286, 2138307, 2137603 and 2138296. We also used resources of the National Energy Research Scientific Computing Center, supported by the Office of Science of the U.S. DoE under Contract No. DE-AC02-5CH11231. DB acknowledges the support of Way Klingler Sabbatical Fellowship. FL and FT acknowledge the Regional Council of Brittany for supporting this study. FL and FT acknowledge the support from the CEA/GENCI (Grand Equipement National de Calcul Intensif) for awarding access to the TGCC (Très Grand Centre de Calcul) Joliot Curie/IRENE supercomputer within the A0110413001 project. Numerical calculations were also performed on the SCaPhyR (Serveur de Calcul pour la Physique Rennaise) cluster. RD and EQS are supported by the United States Department of Energy (DOE), Grant No. DE-SC0025420.
\end{acknowledgements}
%

\bibliographystyle{aa}
\bibliography{references}







   
  



\begin{appendix}




\onecolumn
\section{\label{AppA} Collisional excitation by H$_2$}

\subsection{Interaction Potential}

\begin{figure}[h]
 \centering
 \includegraphics[width=0.45\textwidth]{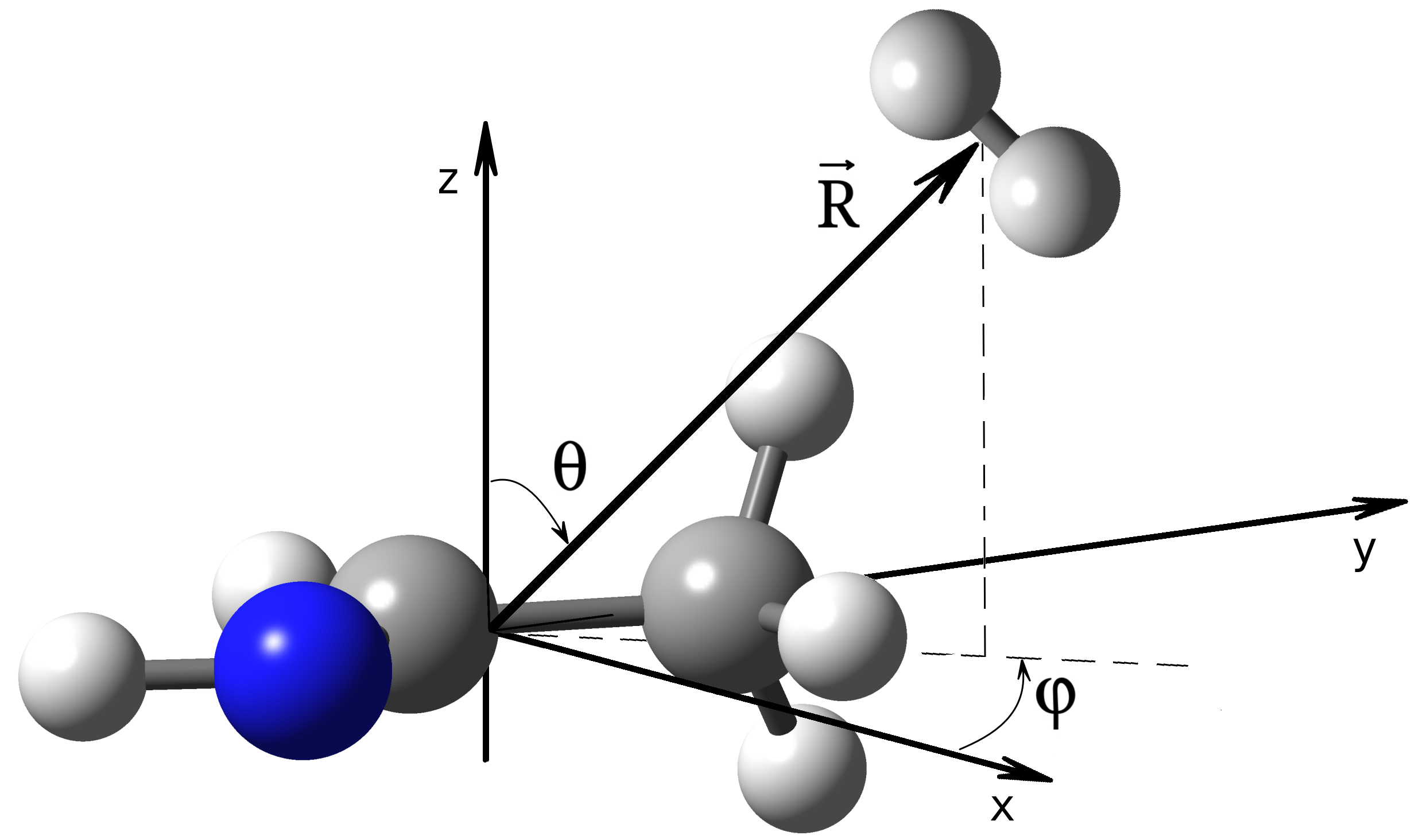}
 \caption{Coordinates used to describe the 3D interaction of the ethanimine molecular isomers ($E$ and $Z$, $E$-isomer shown here) with \textit{para}-H$_2$ in $J=0$. See the text for details.}
  \label{SR}
\end{figure}

Two new potential energy surfaces were constructed for this study, representing each of ethanimine's isomers interacting with \textit{para}-H$_2$. 
Extending our previous study of He-atom interactions with ethanimine (Paper I),
the geometries for the two isomers ($E$ and $Z$) were held rigid using previously reported vibrationally averaged geometric parameters, which are consistent with the observed rotational constants.
Although both ethanimine isomers exhibit large-amplitude motions associated with the internal rotation of the methyl group, the corresponding lowest excited states lie above 150 cm$^{-1}$~\citep{melli2018rotational}. At the low temperatures considered here (5-100\,K), these states are only weakly populated and are not expected to significantly affect the excitation balance, thus supporting the use of the rigid-rotor approximation.
With each ethanimine isomer oriented in its principal axis frame (which includes having its center of mass at the origin and the Cs molecular symmetry plane in the $x$-$y$ plane), interactions with \textit{para}-H$_2$ in $J=0$ can be described by just three intermolecular coordinates: $R$, $\theta$, and $\varphi$. The reduction from five dimensions to three (relevant to other states of H$_2$) is due to the spherical nature of \textit{para}-H$_2$ in $J=0$, where the rotational state's probability density is a constant with no angular dependence. 
$R$ represents the distance between the centers-of-mass of the ethanimine and hydrogen molecules, while $\theta$ and $\varphi$ represent the spherical angles, as illustrated in Figure~\ref{SR} for the $E$-isomer. 

The PESs were constructed using an automated interpolating moving least squares methodology, as implemented in the software package AUTOSURF~\citep{Dawes2018,Quintas2019}, using points computed at the CCSD(T)-F12b/VTZ-F12 level of theory. For each point in the 3D data set, a spherical average over the orientation of H$_2$ was obtained using a 5-point numerical formula~\citep{wernli2007rotational}. Thus each data point in the fitted PES incorporates the results of 5 separate \textit{ab initio} calculations spanning all the required poses of H$_2$.  
All \textit{ab initio} calculations were performed using the Molpro electronic structure code package~\citep{Werner2012-molpro}.
For both ethanimine isomers, the shortest intermonomer center-of-mass distance considered is $R = 2.3$~\AA, and the repulsive part of the PES was further restricted by excluding regions with energies higher than $8$~kcal/mol ($\sim 2\,800$~cm$^{-1}$) above the separated monomers asymptote. 
The \textit{ab initio} data coverage in the fitted PES extended to $R=18.0$~{\AA}, while the zero of energy was set at infinite separation of the monomers.

\begin{figure}[h]
 \centering
 \includegraphics[width=0.45\textwidth]{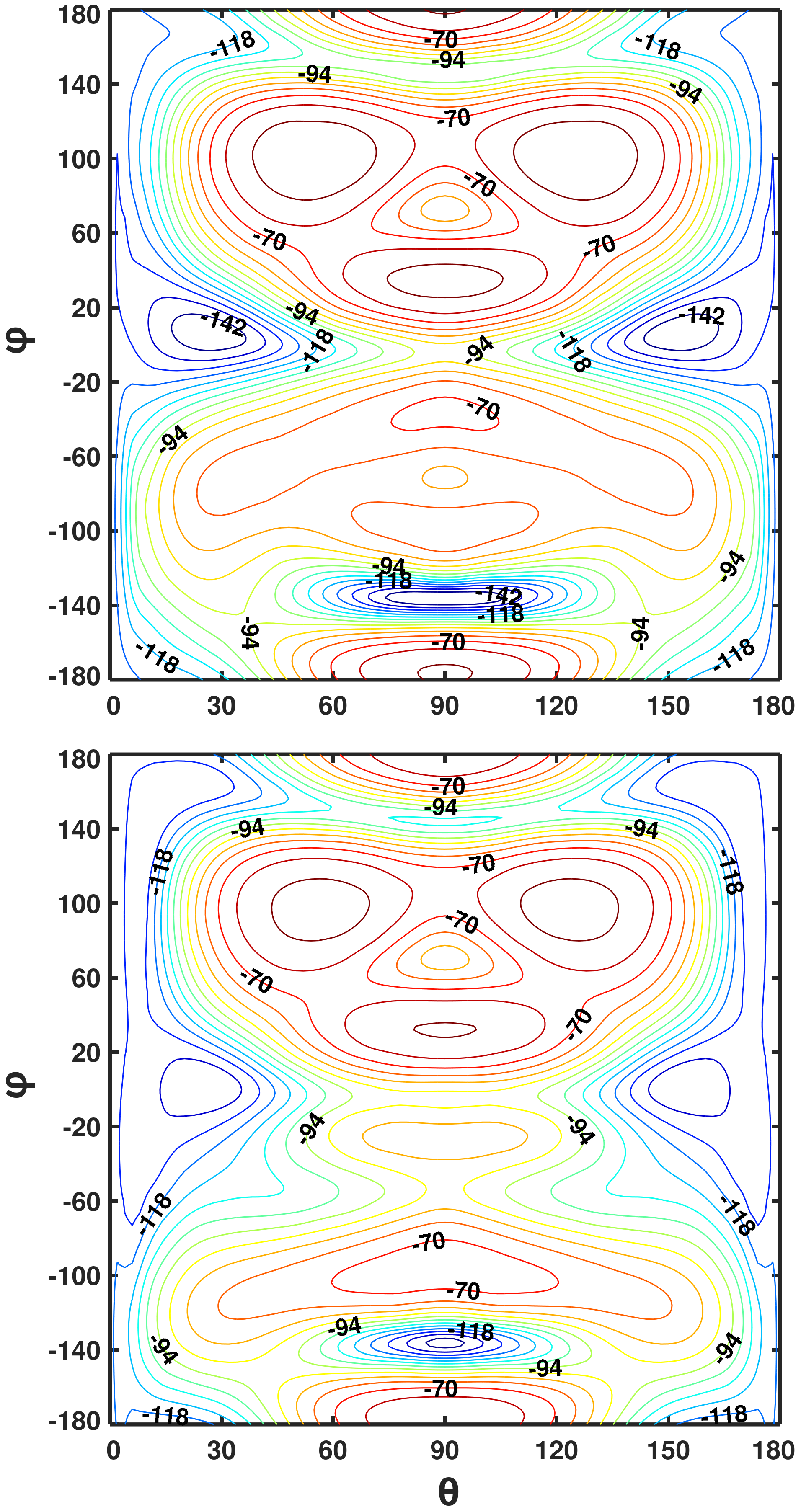}
 \caption{$R$-optimized contour plot of the PES for each isomer (upper: $E$-ethanimine + \textit{para}-H$_2$, lower: $Z$-ethanimine + \textit{para}-H$_2$) as a function of the spherical angles $\theta$ and $\varphi$. See the text for details.}
  \label{Ropt}
\end{figure}

\begin{table}[b]
\caption{\label{TB1} Geometric parameters and well depths for the four minima of the \textit{para}-H$_2$ + ethanimine molecule complex. Units are \AA\,, degrees, and cm$^{-1}$.}
	\centering
	\begin{tabular}{lrrrr}
     \multicolumn{5}{c}{Ethanimine $E$ + \textit{para}-H$_2$} \\
    \hline
           & $R$   & $\theta$ & $\varphi$  & $V$ \\
	\hline
	Min. 1 & $3.315$ & $27.0$ &   $6.1 $ & $-146.6$ \\
	Min. 2 & $3.974$ & $90.0$ &  $145.7$ & $-104.8$ \\
	Min. 3 & $4.667$ & $90.0$ &   $73.0$ & $-85.2$ \\
	Min. 4 & $4.589$ & $90.0$ &  $-71.5$ & $-83.7$ \\
	Min. 5 & $3.867$ & $90.0$ & $-136.3$ & $-148.6$ \\
    \hline
     \multicolumn{5}{c}{} \\
     \multicolumn{5}{c}{Ethanimine $Z$ + \textit{para}-H$_2$} \\
    \hline
           & $R$   & $\theta$ & $\varphi$  & $V$ \\
	\hline
	Min. 1 & $3.328$ & $25.9$ &    $2.0$ & $-135.2$ \\
	Min. 2 & $3.954$ & $90.0$ &  $143.8$ & $-106.9$ \\
	Min. 3 & $4.683$ & $90.0$ &  $ 70.6$ &  $-85.0$ \\
	Min. 4 & $3.959$ & $90.0$ &   $-5.1$ &  $-90.1$ \\
	Min. 5 & $3.832$ & $90.0$ & $-136.3$ & $-138.9$ \\
	\end{tabular}
\end{table} 

For the $Z$-isomer, the global root-mean-squared error (rmse) of the fitted PES is $1.2 $~cm$^{-1}$ (excluding the long range, where the fitting error is extremely small), and the total number of automatically generated symmetry-unique points needed to reach that target was 2784 (the estimated error is $0.1$~cm$^{-1}$ for energies below the asymptote); while for the $E$-isomer, the global rmse is $1.5$~cm$^{-1}$ ($0.1$~cm$^{-1}$ for energies below the asymptote), with 2712 \textit{ab initio} points. 
To guide the placement of high-level data, a lower-level guide surface (double zeta basis) was first constructed using 2836 symmetry-unique points for the $Z$-isomer (2984 points for the $E$-isomer), distributed using a Sobol sequence~\citep{sobol1976uniformly} biased to sample the short-range region more densely. 
To represent the long range (out to arbitrary distances), the PES switches smoothly to an analytic expression representing electrostatic, induction, and dispersion interactions between the fragments (truncated here at 8th order). 
The long-range representation was produced using the recently released LRF software~\citep{batista2026long}.
Fortran codings of the two global PESs are available from the authors upon request.

Figure~\ref{Ropt} shows a representation of the PES (denoted $R$-optimized) for each isomer, as a function of the angles $\theta$ and $\varphi$. 
The plots describe the complete angular ranges, relaxing the energy along the intermonomer distance coordinate $R$ for each pair of angles. 
As can be seen in the figure, both PESs have qualitatively similar topography, characterized by five symmetry-unique minima. The energies and geometric parameters of the minima are given in Table~\ref{TB1}. The nature of the minima is similar to those of the ethanimine-He systems reported previously, but the wells here are roughly three times deeper. In terms of the impact on dynamics, deeper wells support more bound states, but this is offset by the lighter mass of H$_2$ compared to helium.

\subsection{Scattering Calculations}

Using the new interaction potential, we performed MQCT simulations of inelastic scattering to obtain state-to-state rate coefficients of $E$- and $Z$-ethanimine in collisions with \textit{para}-H$_2$. Since in the interaction potential the H$_2$ molecule is treated as a pseudo-atom restricted to its lowest rotational state, the potential $V$ can be represented by the same analytic expansion that was used in Paper I for the ethanimine + He system: 
\begin{equation}
V(R, \alpha, \beta, \gamma)=\sum_{\lambda \mu} v_{\lambda \mu}(R) \tau_{\lambda \mu}(\alpha=0, \beta, \gamma)\,,
\end{equation}
where 
\begin{equation}
\tau_{\lambda \mu}(\alpha=0, \beta, \gamma)=\frac{1}{1+\delta_{\mu 0}}\left[(-1)^\mu \mathrm{Y}_\lambda^{+\mu}(\beta, \gamma)+\mathrm{Y}_\lambda^{-\mu}(\beta, \gamma)\right]\,.
\end{equation}
Here, ($\alpha, \beta, \gamma$) is a set of Euler angles used to position the ethanimine molecule in space, while $R$ is the distance between the centers of mass of ethanimine and H$_2$. The range of indexes of analytic functions $\tau_{\lambda \mu}$ and coefficients $\nu_{\lambda \mu}$ runs through both even and odd values of $\mu$ $(0\leq \mu \leq \lambda)$ with $\lambda \geq 0$. The expansion was truncated to 138 terms, including all $\mu$ values for all $\lambda$ up to $\lambda=15$, plus two more $\lambda=16$ terms with $\mu=0$ and 1. It was established in Paper I that the accuracy of this expansion is sufficient to describe the ethanimine molecule.
To verify that the collision dynamics by H$_2$ does not require additional terms in the PES expansion, we compared, at 600\,cm$^{-1}$, the cross sections from the ground state to all states below 50\,cm$^{-1}$ with those obtained using an extended expansion including 190 terms (up to $\lambda=18$ and $\mu=18$). This comparison validated the use of the reduced 138 term expansion, showing absolute relative differences of 0.1\% for the dominant transitions, below 2\% for transitions of intermediate magnitude, and always below 7\% for the weakest transitions.

\begin{figure*}[h!]
   \centering
   \includegraphics[scale=1.00]{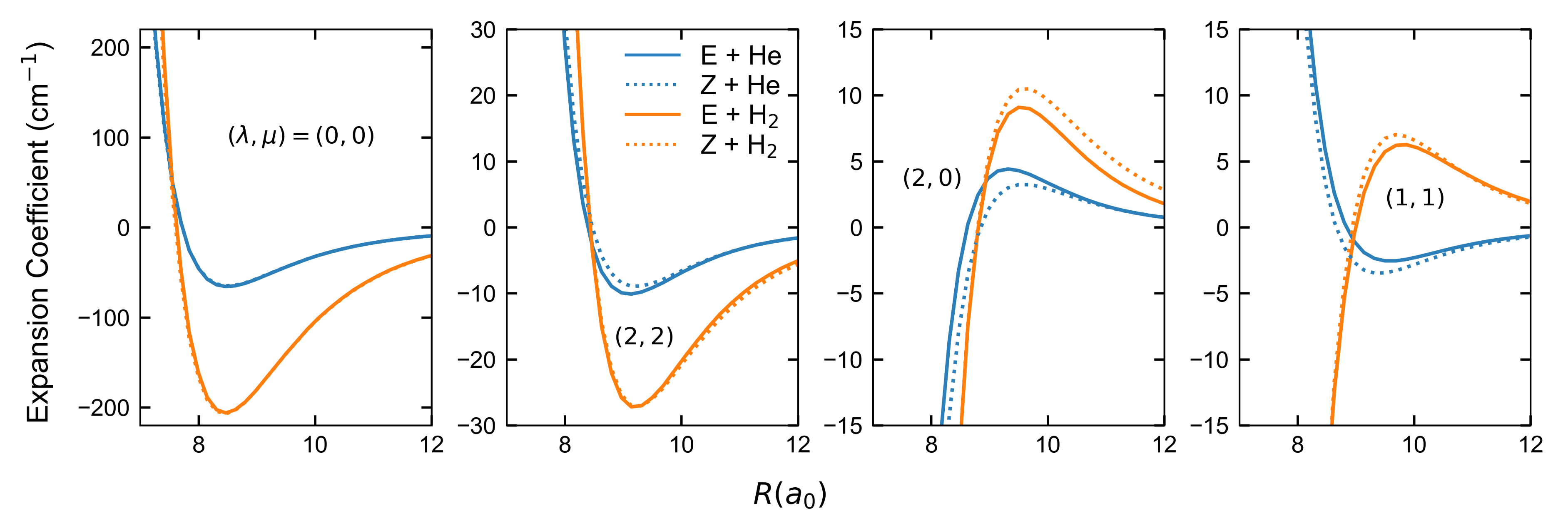}
      \caption{Radial dependence of dominant expansion coefficients $\nu_{\lambda \mu} (R)$ for the interaction of $E$- and $Z$-isomers of ethanimine with the He atom and H$_2$ molecule treated as a pseudo-atom.}
         \label{fig4}
\end{figure*}

Several dominant expansion terms for the interaction of two isomers of ethanimine with He and H$_2$ (four combinations) are presented in Fig.~\ref{fig4}. By comparing the isotropic terms $(\lambda, \mu)=(0,0)$ for the four cases one can conclude that their difference is negligible for the two isomers of ethanimine but is significant for H$_2$ \textit{vs.} He. In the case of H$_2$ the attractive well is about four times deeper compared to He. 
Moreover, in the case of H$_2$ the attractive well and the repulsive wall are both shifted to larger values of $R$, which reflects a larger effective size of the H$_2$ molecule compared to the He atom. By comparing anisotropic expansion terms for the four cases, we conclude that, despite the fact that H$_2$ is treated as a pseudo-atom, its interaction with ethanimine is still much more anisotropic than that of He. Namely, the leading quadrupole term $(\lambda, \mu)=(2,2)$ is almost three times larger in the case of H$_2$, while the other two dominant expansion terms, 
the quadrupole $(2,0)$ and dipole $(1,1)$ terms,
are both about two times larger in the case of H$_2$. Again, the interaction is shifted to noticeably larger values of $R$ in the case of H$_2$, reflecting its larger effective size. The difference between $E$- and $Z$-isomers of ethanimine is small but visible in the case of expansion terms with $(\lambda, \mu)=(2,2), (2,0)$ and $(1,1)$.

In the scattering calculations for ethanimine + H$_2$ we used the same convergence parameters and rotational basis sets that were used in Paper I for ethanimine + He. The only difference is due to the fact that H$_2$ is twice lighter compared to He, which leads to smaller values of maximum orbital angular momentum $l_\text{max}$ required for convergence. For example, for calculations at collision energies $U= 40,100$ and 600~cm$^{-1}$ we used $l_\text{max}=26,44$ and 109, respectively. The corresponding values of maximum impact parameter $b_\text{max}$  were 20~$a_0$   for both  $U= 40$ and 100~cm$^{-1}$ and 15~$a_0$  for $U= 600$~cm$^{-1}$. These MQCT calculations gave us reasonably well-converged cross sections for 3655 transitions between the individual rotational states of the $E$-isomer, and 3828 transitions of the $Z$-isomer (at energies up to 50~cm$^{-1}$). Both quenching and excitation cross sections were computed for each transition, and their weighted average was used to obtain rate coefficients by integration over thermal distribution of collision energies, according to Eqs.~(1-2) from \cite{joy2024rate}.


\end{appendix}

\end{document}